\documentclass[12pt]{article}
\usepackage{epsfig,cite}
\usepackage {amsmath}
\usepackage{amssymb}
\usepackage{ulem}
\include{epsf}
\newcount\timecount
\newcount\hours \newcount\minutes  \newcount\temp \newcount\pmhours
\hours = \time \divide\hours by 60 \temp = \hours \multiply\temp by
60 \minutes = \time \advance\minutes by -\temp
\def\hour{\the\hours}
\def\minute{\ifnum\minutes<10 0\the\minutes
            \else\the\minutes\fi}
\def\clock{
\ifnum\hours=0 12:\minute\ AM \else\ifnum\hours<12 \hour:\minute\ AM
      \else\ifnum\hours=12 12:\minute\ PM
            \else\ifnum\hours>12
                 \pmhours=\hours
                 \advance\pmhours by -12
                 \the\pmhours:\minute\ PM
                 \fi
            \fi
      \fi
\fi }

\def\monthname{\relax\ifcase\month 0/\or January\or February\or
   March\or April\or May\or June\or July\or August\or September\or
   October\or November\or December\else\number\month/\fi}

\def\bold#1{\setbox0=\hbox{$#1$}%
     \kern-.025em\copy0\kern-\wd0
     \kern.05em\copy0\kern-\wd0
     \kern-.025em\raise.0433em\box0 }

\def\beq{\begin{equation}}
\def\eeq{\end{equation}}

\def\ga{\mathrel{\raise.3ex\hbox{$>$\kern-.75em\lower1ex\hbox{$\sim$}}}}
\def\la{\mathrel{\raise.3ex\hbox{$<$\kern-.75em\lower1ex\hbox{$\sim$}}}}
\def\gev{{\rm \, Ge\kern-0.125em V}}
\def\tev{{\rm \, Te\kern-0.125em V}}
\def\gyr{{\rm \, G\kern-0.125em yr}}

\def\gappeq{\mathrel{\rlap {\raise.5ex\hbox{$>$}}
{\lower.5ex\hbox{$\sim$}}}}
\def\lappeq{\mathrel{\rlap{\raise.5ex\hbox{$<$}}
{\lower.5ex\hbox{$\sim$}}}}
\def\Toprel#1\over#2{\mathrel{\mathop{#2}\limits^{#1}}}

\def\m12{m_{1\!/2}}

\def\bea{\begin{eqnarray}}
\def\eea{\end{eqnarray}}
\def\beq{\begin{equation}}
\def\eeq{\end{equation}}

\begin{document}
\begin{titlepage}
\pagestyle{empty} \baselineskip=21pt
\rightline{KCL-PH-TH/2012-05, LCTS/2012-02, CERN-PH-TH/2012-010} \vskip 0.7in
\begin{center}
{\large{\bf Prospective Constraints on Neutrino Masses from a 
Core-Collapse Supernova}}
\end{center}
\begin{center}
\vskip 0.2in {\bf John~Ellis}$^{1,2}$, {\bf Hans-Thomas~Janka}$^{3}$,
{\bf Nikolaos~E.~Mavromatos}$^{1,2}$, {\bf Alexander~S.~Sakharov}$^{2,4}$ 
and {\bf
Edward~K.~G.~Sarkisyan}$^{2,5}$ \vskip 0.1in

{\it
$^1${Theoretical Particle Physics and Cosmology Group, Department of Physics, \\
King's College London, Strand, London WC2R 2LS, UK}\\
$^2${Physics Department, CERN, CH-1211 Geneva 23, 
Switzerland}\\
$^3${Max-Planck-Institut f\"ur Astrophysik, Karl-Schwarzschild-Str. 1, \\
D-85748 Garching, Germany}\\
$^4${Department of Physics, Wayne State University, Detroit, MI 48202,USA}\\
$^5${Department of Physics, University of Texas at Arlington, Arlington, 
TX 76019, USA}\\
}

\vskip 0.2in {\bf Abstract}
\end{center}
\baselineskip=18pt \noindent

We discuss the prospects for improved upper limits on neutrino masses
that may be provided by a core-collapse supernova explosion in our galaxy,
if it exhibits time variations in the neutrino emissions on the scale of a
few milliseconds as suggested by recent two-dimensional simulations.
Analyzing simulations of such neutrino emissions using the wavelet
technique adopted in~\cite{EJMSS}, we find that an upper limit $m_{\nu}
\sim 0.14$~eV could be established at the 95~\% confidence level if the
time variations in emissions were to be preserved during neutrino
propagation to the Earth.

\vspace*{0.2cm}

\vfill \leftline{February 2012}
\end{titlepage}

\section{Introduction}

The observation of a neutrino pulse from supernova SN~1987a has provided
many of the most sensitive probes of neutrino
properties~\cite{neutrinos,imbsn1987a}, notably including interesting
upper limits on neutrino masses. Initial estimates yielded upper limits
$m_\nu \sim {\cal O}(10)$~eV~\cite{oldmnu}, but a recent 
analysis~\cite{Vissani} has 
derived the
stronger upper limit $m_\nu < 5.8$~eV, thanks to improved understanding
of one-dimensional (spherically-symmetric) neutrino emission models.


 Based on a recently developed new generation of two-dimensional 
 simulations (axially-symmetric with polar grid)
 of core-collapse supernovae~\cite{2D}, 
 we have reported recently~\cite{EJMSS} on the sensitivity to
Lorentz-violating effects in neutrino propagation that could be obtained
if the time variations are observed in the neutrino emissions from a
future core-collapse supernova in our galaxy. It has long been appreciated
that SN 1987a provides the most stringent upper limits on an
energy-independent deviation of the neutrino velocity $\delta v$ from that
of light~\cite{olddeltav}, and also strong upper limits on possible
dependences $\delta v \sim E, E^2$~\cite{Harries}. These limits have
recently attracted increased attention, as they constrain significantly
models for the OPERA report~\cite{OPERA} of superluminal neutrino
propagation~\cite{AEM,Giacomo}. It was also shown in~\cite{Harries} that
these limits could be improved if another galactic supernova were to be
observed.



In the present paper we report on a study of the sensitivity to neutrino
mass that would be provided if the time variations found in these
two-dimensional simulations were indeed to be observed.
  This prospective sensitivity is very competitive with other constraints
on neutrino masses, and provides additional motivation (if it is needed) for
further validation of the results of two-dimensional core-collapse
supernova simulations~\cite{dataSim,dataSim2,2D}, particularly via the
development of robust three-dimensional simulations~\cite{3D-1,3D-2}.
Such simulations could be expected to modify the results presented here,
with a tendency to reduce the observability of any time structures in the
neutrino signal. 
 We note also that we make other 
 assumptions that 
 are
 on the optimistic side, e.g., we use the signal from one radial ray,
rather than a full hemisphere, we follow~\cite{2D} in using a relatively 
soft equation of state~\cite{lattimer}, 
and we neglect neutrino oscillations, which are difficult to quantify with generality.


\section{Two-Dimensional Simulation of a Core-Collapse Supernova}


As discussed in~\cite{2D}, the neutrino emission during the post-bounce
accretion phase in the two-dimensional simulation (unlike its
one-dimensional counterpart) exhibits rapid time-variability because of
anisotropic mass flows in the accretion layer around the newly-formed
neutron star. 
These flows lead to large-scale, non-radial mass motions in the layer
between the proto-neutron star surface and the accretion shock, creating
hot spots that can produce transiently in preferred directions neutrino
radiation that is more luminous and with a harder spectrum.
These temporal variations in the luminosities and mean energies are
expected to persist during the hundreds of milliseconds length of the
accretion phase. Such variations could yield fractional changes of 10\%
or more in the emissions of electron neutrinos and antineutrinos
during the most violent phases of core activity in two-dimensional models
with no or only slow rotation~\cite{2D,dataSim2}.
Smaller effects are expected for muon and tau neutrinos, because
lower fractions of them are produced in the outer layers of the
proto-neutron star where asymmetric accretion causes the largest
perturbations. The fluctuating neutrino emission has been shown~\cite{2D} to be detectable in the
IceCube detector~\cite{ice} in the case of a neutrino burst from a  future Galactic
supernova, with typical frequencies between several tens of Hz and 
roughly 200~Hz~\cite{2D}.
  Uncertainties in these predictions include the possibility of a 
  stiffer
  nuclear equation of state, the neutrino transport description 
  that is
  used, and (most importantly) the two-dimensional nature of the 
  simulation.  

We base our analysis here on the the maximal effects to be expected within
the mature two-dimensional models currently available. We therefore
consider emission of electron antineutrinos from the (north-)pole as
predicted by the 15\,$M_\odot$ simulation with the relatively soft
equation of state of Lattimer \& Swesty~\cite{lattimer}, as presented in~\cite{2D}, with no
averaging over a wider range of latitudes. Possible flavor conversions
between electron antineutrinos and other antineutrino flavours are ignored.

\section{Wavelet Analysis Technique}

We use a wavelet transform technique (see~\cite{malat} for a review and~\cite{EJMSS}
for a more detailed description of the approach used here) to analyze the neutrino time series
generated by the simulated supernova explosion. 
 
\begin{figure}[htb]
\begin{center}
\includegraphics[width=0.8\textwidth]{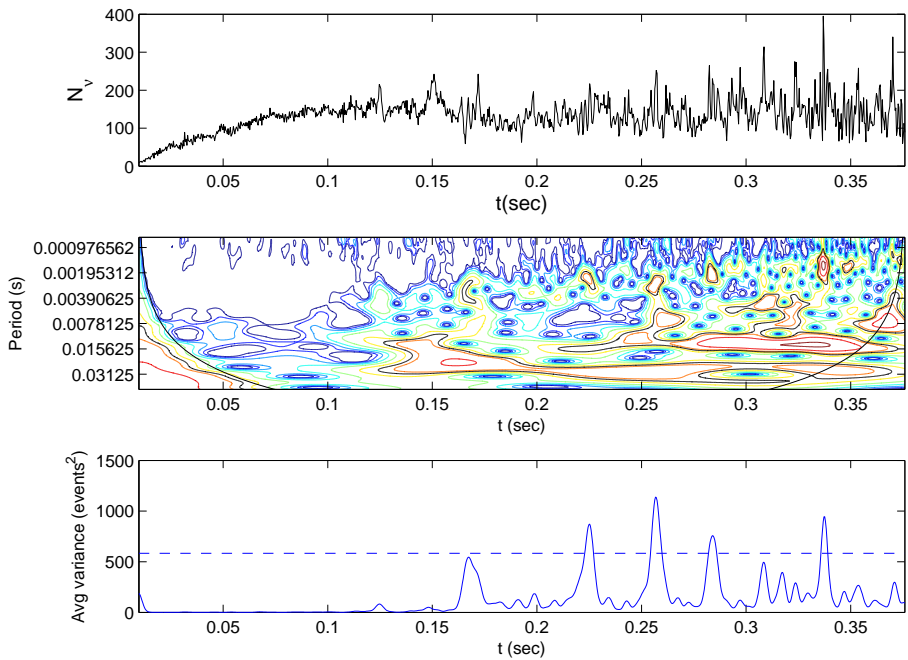}
\end{center}
\caption{\it 
{\underline {\it Top panel:}}
The time series of the neutrino emission from the
two-dimensional
simulation of 
a core-collapse supernova found in~\protect\cite{2D}. 
The time
profile is sampled in 1024 ($2^{10}$) bins. 
{\underline {Middle panel:}} 
The local
wavelet power spectrum of the neutrino emission time series, obtained 
using the Morlet
wavelet function~(\protect\ref{morletWave}) normalized by $1/\sigma^2$~\protect\cite{EJMSS}.
The vertical axis is the
Fourier period (in seconds), and the horizontal axis is the time of the neutrino emission. 
The red contours enclose regions that differ from white noise at greater than the 95\% 
   CL. 
The cone of influence, where edge effects
become important, is indicated by the concave solid lines at the edges of the support of the
signal. 
 {\underline {Bottom panel:}}
 The average power in the 0.002 - 
0.003~s band. The dashed
 line shows the 95\% CL significance. 
 }
\label{fig:w_img}
\end{figure}

We use the Morlet wavelet, which is non-orthogonal,
complex, and contains a number of oscillations sufficient to detect 
narrow features of the power spectrum. We recall that
it consists of a plane wave modulated by a Gaussian function in a variable $\eta$:
\begin{equation}\label{morletWave}
\psi_0(\eta )=\pi^{-1/4}e^{i\omega_0\eta}e^{-\eta^2/2},
\end{equation} 
where $\omega_0$ is a dimensionless frequency.
The continuous wavelet transform of a discrete sequence $x_n$ is defined as the
convolution of $x_n$ with a scaled and translated version of 
$\psi_0(\eta )$:~\footnote{The subscript 0 on $\psi$ has been dropped, in 
order to indicate that $\psi$ has also been normalized (see later).}
\begin{equation}\label{transform}
W_n(s)=\Sigma_{n'=0}^{N-1}x_{n'}\psi^*\left[\frac{(n'-n)\delta t}{s}\right] .
\end{equation}
   In our analysis, the $x_n$ are obtained from a number of independent
statistical realizations of the neutrino signal calculated in~\cite{2D},
as discussed in more detail in the last paragraph of section 2.3 
 of~\cite{EJMSS}.
By varying the wavelet scale $s$ and translating along the localized time 
index $n$, one can construct a picture showing both the amplitude of any
features versus the scale and how this amplitude varies with
time. Although it is possible
to calculate the wavelet transform using (\ref{transform}), it is convenient
and faster to perform the calculations in Fourier space.
According to the convolution theorem, the wavelet transform is the Fourier
transform of the product:
\beq\label{dft2}
W_n(s)=\Sigma_{k=0}^{N-1}\hat x_k\psi^*(s\omega_k)e^{i\omega_kn\delta t},
\eeq
where $\omega_k=+\frac{2\pi k}{N\delta t}$ 
 and $-\frac{2\pi k}{N\delta t}$
for $k\le\frac{N}{2}$ and
$k>\frac{N}{2}$, respectively.
After suitable normalization~\cite{EJMSS}, the expectation value of
$|W_n(s)|^2$ for a white-noise process is $\sigma^2$ for all $n$ and 
$s$. 
We choose discrete scales related by powers of two:
\beq\label{scales1}
s_j=2^{j\: \delta j}s_0,\qquad j=0,1,\dots ,J, 
\qquad J={1\over \delta j}\,\log_2\left(\frac{N\delta t}{s_0}\right),
\eeq
where $s_0$ is
the smallest resolvable scale and $J$ determines the largest 
scale. In the middle panel of Fig.~\ref{fig:w_img} we 
use: 
$N=1024$, 
$\delta t=1.785\cdot 10^{-4}$~s, $s_0=2\delta t$, $\delta j=0.125$ and $J=48$.
In our subsequent analysis, we determine significance levels for the
wavelet spectra with reference to a Gaussian white-noise background spectrum.

\section{Prospective Limit on the Neutrino Mass}

We display in the top panel of Fig.~\ref{fig:w_img} the neutrino time series found in~\cite{2D}, 
summing over all the produced neutrino energies.
It has structures on time scales below a hundredth of a second that lie 
beyond the fluctuations expected from a `featureless' white-noise spectrum.
The middle panel of Fig.~\ref{fig:w_img} shows the normalized wavelet power
spectrum, $|W_n(s)|^2/\sigma^2$, for the time series of the neutrino emission shown in the top panel~\cite{EJMSS}.
The colours represent the significance of the
feature compared to a white-noise spectrum, as measured by the number of $\sigma$ relative to white
noise. Structures in the time series are visible on time scales down to $\sim 2 \times 10^{-3}$~s,
several of which have significance well
above the 95\% confidence level
(CL) for a white-noise spectrum (indicated by red contours). Those with 
time scales 
between 2~ms  and 3~ms can be seen in 
the bottom panel of Fig.~\ref{fig:w_img}. 
We focus on these, rather than structures on longer time scales,
aiming at the best possible time resolution~\footnote{See, however, the caveats in~\cite{EJMSS}.}.

We investigate here how these structures would be smeared out by the
effect of a neutrino mass on its velocity $v_{\nu}$:
 \beq
 \label{mv}
\frac{v_{\nu}}{c}=1 - \left(\frac{E}{m_{\nu}}\right)^{2}.
\eeq
The neutrino data collected from a supernova
explosion will consist of a list of individual neutrino events with
measured energies $E_i$ and arrival times $t_i$, whereas
the results of the simulation in~\cite{2D} are
presented as a set of energy fluxes within time periods of durations $\simeq 3 - 5~{\rm ms}$.
Each of these fluxes may be treated as a black-body spectrum with a specified mean energy. 
We assign statistically to each neutrino in the simulation a specific time of emission and energy,
based on the mean and total energy of the flux in each time period. 
  In order to estimate the sensitivity to the neutrino mass, 
we make 25 statistically independent realizations of the neutrino emission, 
make a wavelet transform of each implementation,
and analyze statistically their sensitivities to $m_\nu$.


Our prospective upper limit on $m_\nu$ is
calculated by requiring that the fine-scale time structures in the wavelet power spectrum 
do not disappear below the 95\% CL of significance for a signal above the white-noise 
power spectrum. Specifically, we apply to each neutrino event an 
energy-dependent time shift 
\beq\label{shiftTime1}
\Delta t= \frac{\tau_{\, m}}{E^{\, 2}},
\eeq
where 
\beq\label{deltaTaum}
\tau_{\,m}=\frac{L  m_{\nu}^2}{c}.
\eeq
We then vary $\tau_{\, m}$ ($m_{\nu}$) 
and follow the evolution of the
signal in the neutrino time series. If there were a non-trivial energy-dependent mass effect during propagation
from the supernova, it could be compensated by choosing the
``correct'' value of the time shift $\tau_{\, m}$, in which case the 
original time structure at the
source is recovered. On the other hand, dispersion at the source itself could not, in general,
be compensated by any choice of $\tau_{\, l}$. Quantitatively, the time 
structure of the supernova signal 
is recovered by maximizing the fraction of the scale-averaged power spectrum above the 95\% 
CL line. In order to calculate a lower limit on $\tau_{\, m}$ in any 
specific model,
we examine the fine-scale time structures that appear
above the 95\% CL in the bottom panel of Fig.~\ref{fig:w_img} and determine
the value of the time-shift parameter (\ref{deltaTaum}) at which the signal above
the 95\% CL disappears.

\begin{figure}[htb]
\begin{center}
\includegraphics[width=0.8\textwidth]{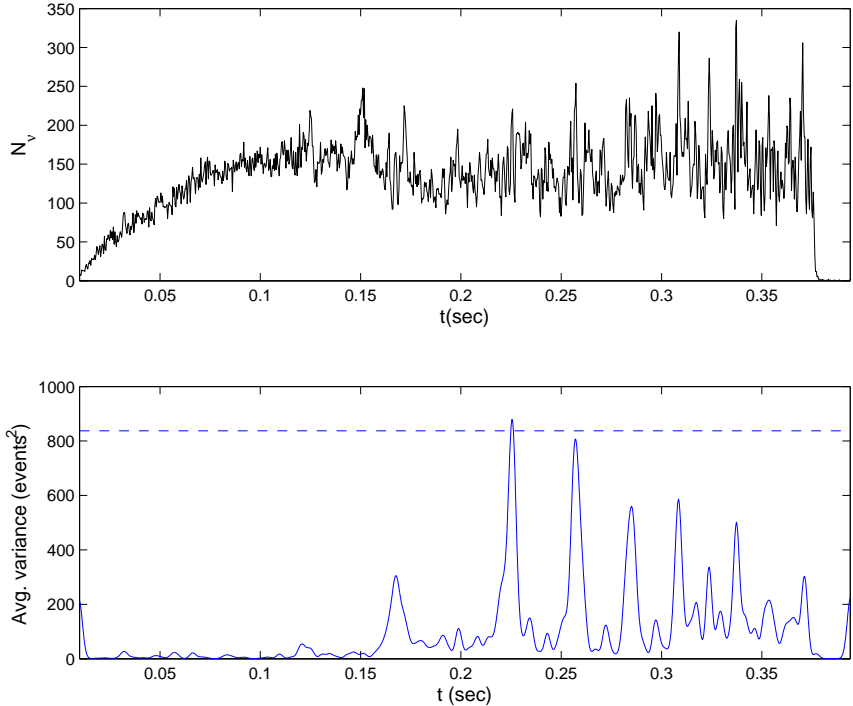}
\end{center}
\caption{\it 
{\underline {Upper panel:}}
The time series of the neutrino emission from the
two-dimensional
simulation of
a core-collapse supernova 
after applying an 
energy-dependent time shift $\tau_{m} = 0.019$~s~$\cdot$~MeV$^2$.
{\underline {Lower  panel:}} 
The strengths of the time-scale structures of the power spectrum averaged 
between
 2 and 3~ms  
 disappear below the 95\% CL of significance after applying this 
time shift.
}
\label{fig:below}
\end{figure}

Fig.~\ref{fig:below} displays the result of one simulation of the effect
of such an energy-dependent refractive index, sampled in 21 bins
corresponding to different time shifts $\tau_{m}$. 
The vertical axis of the lower plot shows
the strength of the emissions in the structures with time scales between
2 and 3~ms, applying an
energy-dependent time shift $\tau_{m} = 0.019$~s~$\cdot$~MeV$^2$. Looking 
at 
the
structures that occur between 0.22 and 0.34~s after the start, we see that
their significant parts (those above the 95\% fluctuation level for
white-noise background
(as seen in Fig.~\ref{fig:w_img})
disappear for time delays $\tau_m = 0.019$~s~$\cdot$~MeV$^2$ and above, 
corresponding to 
$m_{\nu} > 0.14$~eV if a supernova 
distance $L$ of 10~kpc is assumed. 
This sensitivity is one-and-a-half orders of magnitude more sensitive 
than that found in~\cite{Vissani}, namely
$m_{\nu} < 5.8$~eV, based on a one-dimensional 
simulation of a core-collapse
supernova that did not exhibit the small time-scale structures seen in 
Fig.~\ref{fig:w_img}.

\begin{figure}[htb]
\begin{center}
\includegraphics[width=0.8\textwidth]{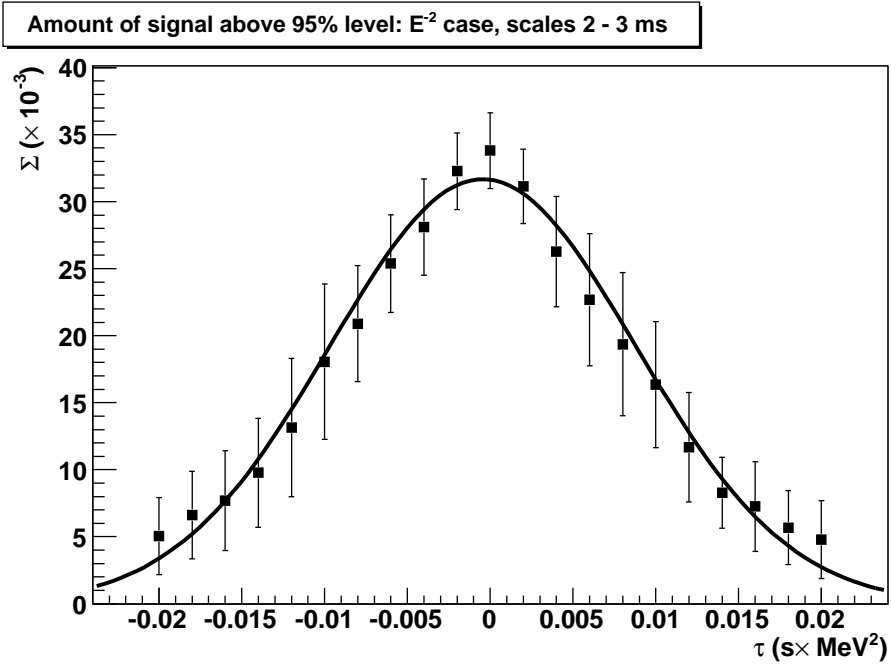}
\end{center}
\caption{\it A Gaussian fit to the amount $\Sigma$ of the short time-scale signal above the 95\%~CL,
calculated for 21 values of the shift parameters $\tau_m$. Each point is obtained as the average 
over 25 realizations of the time-energy assignments of individual neutrinos.}
\label{fig:mFit}
\end{figure}

We have repeated this exercise with 25 different statistical realizations of the
neutrino emission, calculating in each case the amount $\Sigma$ of the total signal above 
95\% CL for different values of $\tau_m$ sampled in 21 bins. The results of these 25 
realizations can be fit quite well by a Gaussian distribution, as seen in
Fig.~\ref{fig:mFit}, which displays our results for the structures
with time scales between 
 2 and 3~ms
 that occur between 0.22 and 0.34~s after the start.
The position of the maximum, which defines the value of  $\tau$ that maximizes the time structures
in the signal and is expected to be zero, is indeed consistent with zero to within a precision 
of $0.001$~s~$\cdot$~MeV$^2$, while the structures are washed out 
at
\beq\label{taumlimit}
\tau_{\, m}= 0.019 \: {\rm s \cdot MeV}^2.
\eeq
Hence, if significant time structures of the type found in the two-dimensional
simulation~\cite{2D} were to be seen in IceCube or a water {\v C}erenkov 
low-energy detector in neutrino data from a core-collapse
supernova at a distance of 10~kpc, one could conclude that
\begin{equation}
m_{\nu} \; < \; 0.14~{\rm eV}.
\label{limitm}
\end{equation}
On the other hand, if no such structures were seen, this could mean either that the 
structure was washed out by the effect of neutrino propagation with
   $m_{\nu}\gtrsim 0.14$~eV (which is still  consistent with the 
cosmological limit $m_{\nu}<0.23$~eV~\cite{cosmo}),
or that the structure found in~\cite{2D,dataSim2} is not 
valid. 

The first possibility could be probed by performing a series of 
analyses of the above type in several different bands of the 
power spectrum~\footnote{We recall that, in order to obtain the greatest 
sensitivity, 
the averaging in Fig.~1 was performed in the $\sim$~2--3~ms band where the 
shortest time-varying structures appear above the 95\% CL. However, there are structures at other scales,
e.g., in the  $\sim$~7--15~ms band.}. 
In practice, in the presence of an apparently structureless original signal, one 
would average systematically over all possible bands with a relatively 
small step of granularity, 
 defined by
 the precision of the 
analysis, 
 while changing $\tau_m$ so as to maximize the amount of the signal 
above the 95\% 
CL for the averaged power in every band included in the scan. 
 As soon as a
value of $\tau_m$ is found 
with a significant fraction of the signal 
 above the 95\% CL line at
$\tau_m \gtrsim 0.019\ {\rm s\cdot MeV^2}$, 
 one 
 could claim evidence for a non-zero neutrino mass. 
 On the other hand, if such a scanning analysis
 delivered a negative result, inferring a lower limit on $m_{\nu}$ would
require strong independent confirmation of the structures found
in~\cite{2D,dataSim2}, in particular by full three-dimensional 
simulations.

We note that the possibility of a time advance is also considered in
Fig.~\ref{fig:mFit}. This would correspond to a neutrino with $m_\nu^2 < 0$,
i.e., a tachyon. As pointed out, e.g., in~\cite{AEM}, the time advance
reported by OPERA~\cite{OPERA} could not be associated with tachyonic neutrinos because,
e.g., this would require an unacceptable time advance $\sim 4$y for neutrinos from
supernova SN1987a~\footnote{There is a more fundamental problem with Lorentz-invariant tachyonic neutrinos,
namely that there are no non-trivial finite-dimensional unitary representations of the Lorentz
group for $m^2 < 0$ that could correspond to spin-1/2 fermions. However, here we restrict
ourselves to phenomenological constraints on tachyonic neutrinos.}. Conversely, SN 1987a provides the strongest available
lower limit on negative $m_\nu^2$. Correspondingly, observation of the short time
structures suggested by~\cite{2D} would establish a much stronger limit on the
possible tachyonic nature of the neutrino: $m_\nu^2 > - 0.02$~eV$^2$, which would also be
significantly stronger than the bound $m_\nu^2 > - 0.11$~eV$^2$ recently derived by combining neutrino constraints
from Big-Bang nucleosynthesis and the cosmic microwave background~\cite{DM}.

\section{Conclusions and Prospects}

We have shown that the existence of structures with short time scales in 
the neutrino
emission from a core-collapse supernova, as suggested by two-dimensional
simulations~\cite{2D,dataSim2}, would open up new prospects for probing
neutrino masses. The sensitivity (\ref{limitm}) extends up to 
one-and-a-half orders of 
magnitude beyond the sensitivity provided by previous analyses 
based on one-dimensional supernova simulations~\cite{Vissani}. This 
sensitivity
is comparable to the $\sim 0.2$~eV obtainable with the KATRIN 
experiment~\cite{KATRIN}, 
and to the potential sensitivity to $m_\nu$ provided by large-scale 
structure surveys
in combination with measurements of the cosmic microwave background 
radiation~\cite{cosmo}.

If such short time structures were not to be seen in emissions from a future core-collapse
supernova, many checks would be necessary before one
could conceivably claim observation of a non-zero neutrino
mass. In particular, it would be necessary to validate the structures predicted by the two-dimensional
core-collapse supernova simulation on which this analysis is based, specifically
in full three-dimensional simulations~\cite{3D-1,3D-2}. 
  It would also be necessary to 
 test the influence of
 some of the other special or 
 simplifying assumptions made here, e.g., 
 the 
use of the signal from one radial ray~\cite{2D},
the use of a relatively soft equation of state~\cite{lattimer}, 
and the neglect of neutrino oscillations.
 On the other hand, convergent indications from supernova observations,
KATRIN~\cite{KATRIN} and astrophysical observations~\cite{hizgalaxies} would substantiate and consolidate
any determination of neutrino mass in the range 0.1 to 0.2~eV.

\section*{Acknowledgements}

The work of J.E. and N.E.M. was supported partly by the London Centre for
Terauniverse Studies (LCTS), using funding from the European Research
Council via the Advanced Investigator Grant 267352.
H.-T.J. acknowledges support by the Deutsche Forschungsgemeinschaft
through the Transregional Collaborative Research Centers SFB/TR~27
``Neutrinos and Beyond'' and SFB/TR~7 ``Gravitational Wave Astronomy'',
and the Cluster of Excellence EXC~153 ``Origin and Structure of the Universe''
({\tt http://www.universe-cluster.de}). The supernova simulations were 
possible by computer time grants at the John von Neumann Institute for
Computing (NIC) in J\"ulich, the H\"ochst\-leistungs\-re\-chen\-zentrum
of the Stuttgart University (HLRS) under grant number SuperN/12758,
the Leib\-niz-Re\-chen\-zentrum M\"unchen, and the RZG in Garching.
We are grateful to Andreas Marek for providing the neutrino data.


\end{document}